\documentclass[twoside,12pt,a4paper]{article}
\usepackage{a4wide}
\usepackage{epsfig}
\usepackage{latexsym}
\begin{document}

\begin{center}
{\Large \bf 
Bootstrap equations in effective theories
}
\end{center}

\begin{center}
A.~Vereshagin\footnote{e-mail:  Alexander.Vereshagin@fi.uib.no} \\
{\em University of Bergen and St.-Petersburg State University} \\
V.~Vereshagin\footnote{e-mail: vvv@av2467.spb.edu},
\,  K.~Semenov-Tian-Shansky\footnote{e-mail: semenov@pdmi.ras.ru} \\
{\em St.-Petersburg State University}
\end{center}
\vskip 2pt
\bigskip
\section{Introduction}
\label{sec-introduction}
\mbox{}

Presently, there are known two different approaches to the problem of
constructing of a relativistic quantum theory suitable to describe the
scattering processes. According to the first --  canonical -- approach
one has to fix a finite set of (fundamental) fields and the Lagrangian
that satisfies certain conditions (locality, hermiticity, symmetry,
renormalizability, etc.). To obtain the quantum version, one has to
carry out the procedure of canonical quantization. This allows one to
construct the Fock space of asymptotic states and to calculate the
Green functions and the
$S$-matrix.

The second approach (first suggested in papers
\cite{anyspin}; we will call it as the
$S$-matrix approach)
is less known and practically was not discussed
in literature  (see however
\cite{Novozhilov}, \cite{RumerFet}
and chapters  2 -- 5 of the monograph
\cite{WeinMono}).
In this approach the structure of the Fock space of asymptotic states
is postulated initially. The field operators are constructed according
to the symmetry properties of the corresponding one-particle states.
The Hamiltonian of the system (built out of these fields) is also
postulated as the operator in the interaction picture.

The elements of the
$S$-matrix
are calculated according to Dyson formula
($f$, $i$ --
the final and initial states respectively):
\begin{equation}
S_{fi} =
\langle f |
T_{\stackrel{\;}{\scriptscriptstyle\!\! W}} exp
\left\{
-i \int\limits_{}^{}
H_{int} dx
\right\}
|\, i \rangle.
\label{1.1}
\end{equation}
The symbol
$
T_{\stackrel{\;}{\scriptscriptstyle\!\! W}}
$
in this formula denotes Wick's (manifestly covariant) T-product. The
non-covariant terms in the Hamiltonian and in propagators are to
be discarded (see
\cite{WeinMono}).
In the case of effective theories only discussed below this does not
lead to any  uncertainties.

Each of these two approaches has its advantages and shortcomings.
However, the comparative analysis is not our goal here. We use the
$S$-matrix
approach just because the canonical quantization of the theories with
Lagrangians containing high (second and higher) powers of the field
time derivatives is the problem the solution of which is presently
unknown. At the same time, the effective theories, which are the main
subject of our study,  contain
{\em all}
the powers of first (and higher) field derivatives by the very
construction.

We use the slightly improved version of the definition of the
effective theory (originally  given in
\cite{WeinET}):
{\em the theory is called as effective if the corresponding
Hamiltonian (in the interaction picture) contains
\underline{all}
the local terms  consistent with the requirements of a given
algebraic symmetry}.
In the operator sense this construction is not very well defined.
However, we are only interested in the
$S$-matrix
elements calculated (only on the mass shell!) with the help of
the expansion based on the formula
(\ref{1.1}).
In the papers
\cite{VV}, \cite{AVVV}
it was shown that for those objects it is possible to formulate simple
correctness conditions for the expressions calculated in the given
order of loop expansion. In the zeroth order (tree graphs) of the
renormalized perturbation theory these conditions lead to reasonable
restrictions on the values of physical parameters of a theory.

The effective theories, by the very construction, show the property
of multiplicative renormalizability (in the case of absence of
anomalies). However, usually they are not considered in the
textbooks on renormalization theory  (see for example
\cite{Collins}).
The reason is that to fix the physical content of such a theory one
needs to impose an infinite set of renormalization prescriptions
(conditions). It is absolutely unrealistic if we have no guiding
principle limiting the freedom of this step. Preliminary results of
our research (see
\cite{VV}, \cite{AVVV})
show that the analyticity-type restrictions could play the role of
such a principle. Moreover, they seem to be natural from the point of
view of correctness of perturbative scheme. For example, one could
obtain the bootstrap conditions (widely discussed in connection with
dual models -- see
\cite{Frampton})
from the properly formulated requirements of meromorphy and
polynomial boundedness of the tree-level amplitudes constructed in
the framework of renormalized perturbation theory.

In this paper we illustrate the techniques of derivation of the
bootstrap equations from the condition that the function of two
complex variables is meromorphic and polynomially bounded in
different
domains. However, before starting the analysis of the examples we
would like to outline the reasons why these conditions
(automatically fulfilled in conventional field theories)  turn out
to be fruitful in the case of effective theory. Three following
sections serve for this purpose.

\section{Preliminary notes}
\label{sec-preliminaries}
\mbox{}

First of all we remind the meaning of some notions and terms used
below. We specify the definitions given in
\cite{WeinMono}
and introduce the notion of minimal parametrization of the effective
theory. We use these terms because they are suitable for the work with
on-shell matrix elements in terms of complex analysis.

Let us emphasize that in what follows it is assumed that the masses of
the particles with spin
$J > 1/2$
are nonzero. This assumption is purely a technical one, but at
present we cannot proceed without it. It does not lead to any
limitations when we are describing the strong interaction of
hadrons.

All the combinations of coupling constants that do not appear in the
expressions for the renormalized
$S$-matrix
elements of the
$L$-th
order in loop expansion are called as the
{\em redundant}
parameters of the order
$L$.
These combinations can appear in the expressions for Green functions,
but the corresponding contributions prove to be irrelevant after
renormalization, passing to the mass shell and multiplying by the
wave functions. The example of the redundant parameter is given by the
gauge fixing constant in the renormalizable vector models. Another
example is the wave function renormalization constant
\cite{WeinMono}.

All the independent combinations of coupling constants that
appear in the expressions for renormalized matrix elements of the
$L$-th order of the loop expansion are called as the
{\em essential}
parameters of the order
$L$.

Consider now the
{\em elementary}
(pointlike) vertex with
$n$
legs  carrying the momenta
$p_1, p_2, \ldots , p_n$.
In general, it can be written as follows
\begin{equation}
V(p_1, \ldots , p_n) =
\sum\limits_{a=1}^{M+N} T^{(a)} F_a\ ,
\label{1.2}
\end{equation}
where
$\{ T^{(a)} \}$
is the full set of independent tensor structures
($M$
of them being minimal and
$N$ --
nonminimal; see the definitions below), and
$F_a$
are the functions of invariant kinematical variables
(the total number of those variables is equal to
$3$
when
$n=3$
and
$4n-10$
when
$n > 3$;
we are working in $D=3+1$ dimensions). It is convenient
to choose these variables as follows:
\begin{equation}
[\pi , \nu ] \equiv
[{\pi}_1,...,{\pi}_n; {\nu}_1,...,{\nu}_{3n-10}]\ ,
\label{1.3}
\end{equation}
where
\begin{equation}
{\pi}_i \equiv p^2_i - m^2_i\ .
\label{1.4}
\end{equation}
The concrete choice of the rest
$3n-10$
variables
${\nu}_i$
(independent linear combinations of the momentum scalar products)
is not important for the present.

The vertex
(\ref{1.2})
is the element of the system of Feynman rules of the effective theory
under consideration. It describes the contribution of a term in the
Hamiltonian which is constructed from
$n$
field operators.

It is clear that the functions
$F_a$
polynomially depend on the variables
(\ref{1.3});
the polynomial coefficients are the combinations of coupling
constants. It would be premature to discuss the convergency conditions
for the series of vertices with different number of derivatives. Our
purpose is to obtain the well-defined expressions for the
$S$-matrix
elements of the given order; to obtain them one needs to take
into account not only the vertices of the type
(\ref{1.2}),
but also the contributions of all possible graphs with
$n$
legs. In what follows we would only take care about the correctness of
the expressions that appear as the result of infinite summation of
graphs.

The contribution of the vertex
(\ref{1.2})
to the matrix element describing the process with
$n$
external particles can be obtained by passing to the mass shell
$ {\pi}_i = 0,\ (i=1,...,n) $
and multiplying by the relevant wave functions. Hence, those
combinations of coupling constants which form the coefficients
at nonzero powers of
${\pi}_i$
in the series for
$F_a$,
do not appear in the contribution of the corresponding pointlike
vertex to the
$S$-matrix
element in question. The same is true for the combinations of coupling
constants which appear in expressions for those
$F_a$
that are the coefficients at tensor structures (called as
{\em nonminimal})
resulting in zero after passing to the mass shell and
multiplying by the relevant wave functions.

It is easy to understand that the discussion above is also relevant to
the vertices with arbitrary number of ``bubbles'' (self-closed lines)
and
``tadpoles''
(a line that starts in a vertex and ends by one or several
``bubbles'').
We treat such vertices as pointlike.

Thus, after the vertex
(\ref{1.2})
is put on the mass shell and multiplied by the relevant wave
functions, it takes the form:
\begin{equation}
V(p_1, \ldots , p_n) \stackrel{on\, shell} {\sim}
\sum\limits_{a=1}^{M} T^{(a)} F_a(\pi = 0, \nu)\ .
\label{1.5}
\end{equation}
Here  the set of
$T^{(a)}$ $(a=1,...,M)$
contains only minimal tensor structures and does not contain any
nonminimal ones.

The line
$p_k$
of the vertex
$V(p_1,...,p_n)$
is called as
{\em minimal},
if the explicit form of the expression
(\ref{1.2})
does not change its appearance when the
$k$-th
particle  is put on its mass shell
${\pi}_k = 0$
and the vertex is multiplied by the relevant wave function
$u(p_k)$.

We call the vertex
$V(p_1,...,p_n)$
as
{\em minimal},
if all its lines are minimal. Minimal vertex can always be
presented in the form
(\ref{1.5}).
Finally, we call the propagator of a particle with mass
$m$
and spin
$J$
as {\em minimal},
if its numerator is just a spin sum written in a covariant form
(on the mass shell
$q^2 = m^2$!)
and considered as a function of four independent components of
momentum%
\footnote{It is this point were our suggestion on the absence of
massless particles of higher ($J > 1/2$) spin happens important.}.

Each
$S$-matrix
graph of the effective theory can be rewritten in terms of minimal
elements: vertices (of different orders) and propagators (see
\cite{AV};\
the complete proof will be published later). This means that the
full set of essential parameters includes only masses and those
combinations of coupling constants which define the minimal
parametrization of the vertices of different orders%
\footnote{The explicit expressions of minimal parameters
in terms of the coupling constants certainly depend on the
perturbation order.}.
Thus to obtain the finite
$S$-matrix
it is sufficient to impose only those normalization prescriptions
which are necessary to fix the finite parts of the minimal parameters.
However the results of the papers
\cite{VV}, \cite{AVVV}
show that this set of prescriptions is also excessive. The necessity
to observe the bootstrap restrictions (see below) results in the fact
that only a part of minimal parameters could be treated as
independent.

The renormalization prescriptions should be only imposed on the set
of independent constants of the theory. This means that there are two
(equivalent) methods to impose them. The first method consists in
finding the explicit solution of bootstrap conditions%
\footnote{Such a solution certainly exists:
the example is provided by conventionally renormalizable theories.}.
Following this method one singles out the full set of
{\em independent}
parameters and then imposes the renormalization prescriptions for this
very set. The second method is to impose an arbitrary set of
prescriptions on
{\em all }
the minimal parameters and then to use the bootstrap equations as the
binding limitations on the possible structure of this set. We use the
second method (just because we cannot find the explicit solution of
the bootstrap conditions); so we would like to discuss it in more
detail.

The renormalization prescriptions, irrespectively to their explicit
form, should be satisfied at every fixed order. In particular this is
true with respect to the lowest (tree-level) order. This means that
each relation between the minimal parameters of zeroth order (recall
that in the framework of the renormalized perturbation theory this is
the relation between the physical values of the parameters!) should
be treated as the relation between the prescriptions. In other words,
if there exist some connections between the minimal parameters of
zeroth order (this is exactly the case in the mathematically
reasonable effective theory) then one cannot impose arbitrary
renormalization prescriptions. This statement would be a triviality
if we were discussing the restrictions due to some symmetry (group)
limitations. But in the case of effective theories these restrictions
(bootstrap equations) arise from the certain requirement of
{\em localizability},
discussed in the following section.

\section{Localizable effective theories}
\label{sec-locality}
\mbox{}

In case of ordinary renormalizable theory every term of the loop
expansion of the
$S$-matrix
based on the expression
(\ref{1.1})
is well defined (the regularization is implied). Not much could be
said about the convergency of this expansion, but this does not
create a problem at arbitrary finite order. The situation is quite
different in the case of effective theory. The Hamiltonian contains
the terms with many derivatives (of arbitrary high degree and order)
hence the infinite power series appear in the expressions for matrix
elements already at the tree level. In other words, in this case the
Hamiltonian is not a local operator, and  one has to exercise caution
when working with it. That is why in what follows we will limit
ourselves with the special class of effective theories.

We will only consider the Hamiltonians from the class of
{\em localizable}
ones. One can intuitively  give an idea on localizability considering
the simple example from electrostatics. The interaction Hamiltonian of
point charge with the extended charge is nonlocal. Nevertheless, under
certain conditions (well separated systems) it can be localized
(rewritten in the form of a convergent infinite series of local terms)
with the help of multipole expansion.

The localizability requirements could be briefly formulated in the
implicit form. The explicit formulation would take too much space but
not much illuminate the general idea. The main idea is suggested by
the quasiparticle method
\cite{quasi}
well known in the non-relativistic quantum mechanics.

{\em We call the Hamiltonian as localizable if the tree-level
amplitudes formally obtained from it could be reproduced in the
framework
of well defined tree approximation of a certain extended effective
theory containing auxiliary fields that correspond
to the particles (with masses
$M_i$),
unstable with respect to  decays into the states of  initial theory.}

An important note: in the instability condition
\begin{equation}
M_i \geq m + {\mu}
\label{1.6}
\end{equation}
small letters denote the
{\em physical}
particle masses, i.e. the masses of asymptotic
states of the initial theory  Hamiltonian (see
\cite{WeinMono}).
The quantities
$M_i$
have the meaning of mass parameters of the extended Hamiltonian. They
define the position of the poles of bare propagators. Their treatment,
thus, depends on the renormalization scheme that is being used (see
\cite{Nekrasov}, \cite{Sirlin}).
However the detailed study of this point lies beyond the scope of
this paper. We use the term ``mass'' in both senses, because this
does not lead to misunderstanding.

The words ``well defined'' are to be understood in the sense that the
formal tree level series of the extended theory should be summable in
all the domains of definition. Besides, the tree level series of the
initial theory should converge at least in a small domain
$D$ --
otherwise, the comparison would happen impossible. In this domain the
tree approximations for all the amplitudes describing scattering and
creation of stable particles in both theories should coincide
identically. In other words, in the sector of stable particles the
tree approximation of the extended theory is just the analytic
continuation of the tree approximation of the initial theory. This is
precisely the essence of the extension idea. The condition
(\ref{1.6})
is necessary to ensure the coincidence of asymptotic states in both
theories. In the extended theory the loop corrections lead to
non-stability of the particles described by auxiliary fields and the
{\em asymptotic}
space (the space of stable states) becomes the same as that in the
initial theory. It is well known (see
\cite{Veltman})
that in the theory with unstable particles the
$S$-matrix
constructed  in accordance with the formal Feynman rules turns
out to be a unitary operator on the space of stable states.

Before transforming this philological description of localizable
theories into the definite limitations on the values of coupling
constants we need to make several preliminary remarks.

The requirement of localizability mirrors the natural wish to work
with the series, each term of which is a well defined  function of
momenta. And -- by the very construction -- the general structure of
(\ref{1.1})
provides a guarantee of covariance, causality, unitarity and crossing
symmetry
of the
$S$-matrix.

Up to present the only known tool producing the series with the
desired structure is the formalism of the quantum field theory with
the Hamiltonian  containing the finite number of local interaction
terms.
It is essential that in this approach the tree level amplitude of
every process happens to be a rational function of each pair energy
(with all other variables fixed). It is easy to see that in the case
of the theory containing an
{\em infinite}
number of scalar fields
${\phi}_i$
$(i=1,2,...)$
with masses
$m_i$
and the renormalizable type of interaction
$$
H_{int} =
g_{ijk} {\phi}_i {\phi}_j {\phi}_k +
{\lambda}_{ijkl} {\phi}_i {\phi}_j {\phi}_k {\phi}_l\ ,
$$
no changes are needed  in the general scheme of quantum theory. It
is sufficient that the matrices of coupling constants
$g_{ijk}$,
${\lambda}_{ijkl}$
and the masses
$m^2_i$
satisfy the conditions ensuring the convergence of series at every
given order of loop expansion. When this condition is applied, the
tree-level amplitudes of all processes (not only of those  describing
the scattering and production of stable states) happen to be
meromorphic functions in each pair energy. Aside from this one should
require (see
\cite{AVVV}, \cite{AV})
that these functions are polynomially bounded (in the sense of contour
asymptotics -- see for example
\cite{Shabat})
at zero momentum transfer. This condition is necessary if we would
like to construct the loop graphs by means of closing the external
lines in the corresponding tree graphs, without being anxious about
the order of operations.

In principle, the situation in the initial effective theory will not
differ from that described above if the analytically continued
(from the postulated convergency domain
$D$)
amplitudes turn out to be polynomially bounded meromorphic functions
(not arbitrary;  see below) of each pair energy. The meromorphy
properties make us hope that these functions could be reproduced in
the framework of the extended effective theory, containing the
auxiliary fields with suitable masses. Of course this is not always
possible because not every meromorphic function could be obtained as a
tree level amplitude of some hypothetical field theory. Thus the
requirement of localizability is to be interpreted as the conditions
of the
{\em existence}
of extended effective theory of the most general form. The only
limitation is that the mass spectrum of this latter theory must
satisfy the non-stability condition
(\ref{1.6}).

The localizability requirement leads to certain conditions for the
$S$-matrix elements, we call them as the
{\em analyticity conditions}.
The special term is used because we want to avoid the necessity of
stressing the formal difference between the Hamiltonians of the
extended and initial theories.
{\em The extended theory is introduced just because we have no tool
allowing us to work beyond the frames of Dyson's perturbation
theory.}

In principle, the restrictions on the coupling constants of the
extended theory can be transformed into the desired conditions of
localizability restricting the possible set of couplings in the
initial Hamiltonian. For this it is sufficient to compare the
expansions of the scattering amplitudes calculated in both theories in
the domain
$D$.

\section{Analyticity conditions}
\label{sec-analyticity}
\mbox{}

First of all we need to consider the extended theory and formulate the
analyticity conditions. However, this problem is rather complicated.
Here we are going to take only a first step: we will formulate the
(necessary) analyticity conditions for the amplitudes of binary
processes. This case is relatively simple because the kinematics is
completely described by two independent variables.

It is convenient to introduce three equivalent sets:
\begin{equation}
(x, {\nu}_x)\ , \ \ \ \ \ \ \ \ x=(s,t,u)\ ,
\label{1.7}
\end{equation}
where $(s,t,u)$ stand for conventional Mandelstam variables and
\begin{equation}
{\nu}_s = t-u\ , \ \ \ \ \
{\nu}_t = u-s\ , \ \ \ \ \
{\nu}_u = s-t\ .
\label{1.8}
\end{equation}

The tree-level amplitude
$M(s,t,u)$
of an arbitrary binary process with scalar particles (the
generalization for the case of arbitrary spins does not lead to any
particular difficulties) constructed in accordance with
Feynman rules takes a form of the following formal series:
\begin{equation}
M(s,t,u) = \sum\limits_{i,j,k=0}^{\infty} a_{ijk} s^i t^j u^k +
           \sum\limits_{R_s}^{} \frac{N_s(s,t,u)}{s-M^2_R} +
           \sum\limits_{R_t}^{} \frac{N_t(s,t,u)}{t-M^2_R} +
           \sum\limits_{R_u}^{} \frac{N_u(s,t,u)}{u-M^2_R}\
\label{1.9}.
\end{equation}
The summation should be carried out over all kinematically allowed
resonances
$R_x$ ($M_R$
stand for the corresponding resonance masses) in each channel, and
also (in the first sum) over all four-particle vertices. In turn, the
numerators
$N_x(s,t,u)$
take a form of (formal, maybe infinite) sums
$$
N_x(s,t,u) =
  \sum\limits_{i,j,k = 0}^{\infty} b^{(x)}_{ijk} s^i t^j u^k\ .
$$
Numerical matrices
$a_{ijk}$
and
$b^{(x)}_{ijk}$
are the functions of the coupling constants of the (extended)
Hamiltonian.

The series
(\ref{1.9})
should be summable in order to make sense and to be used for
constructing
the next orders of the loop expansion. The result must be a
meromorphic function in each of the Mandelstam variables. To provide
the possibility of constructing loops and carrying out the
renormalization procedure, this function necessarily%
\footnote{One can convince himself that this is the case when
considering the loop graph as the integral of the product of a tree
graph by a relevant propagator. The integral is to be taken at zero
momentum transfer between the legs that are being closed
\cite{AV}.}
should be polynomially bounded in
${\nu}_x$
at
$x=0$
and, by continuity, in the small vicinity of this value.

With the help of definitions
(\ref{1.8}),
one can rewrite the formal series
(\ref{1.9})
in three different forms:
\begin{equation}
M(s,t,u) =
 \sum\limits_{i=0}^{\infty}{\alpha}^{(s)}_i(s) {\nu}_s^i +
 \sum\limits_{R_t}^{}
      \frac{\rho^{(st)}(s)}{(\sigma - 2M^2_R)-s+{\nu}_s} +
 \sum\limits_{R_u}^{}
      \frac{\rho^{(su)}(s)}{(\sigma - 2M^2_R)-s-{\nu}_s}\ ;
\label{1.10}
\end{equation}
\begin{equation}
M(s,t,u) =
 \sum\limits_{i=0}^{\infty}{\alpha}^{(t)}_i(t) {\nu}_t^i +
 \sum\limits_{R_u}^{}
      \frac{\rho^{(tu)}(t)}{(\sigma - 2M^2_R)-t+{\nu}_t} +
 \sum\limits_{R_s}^{}
      \frac{\rho^{(ts)}(t)}{(\sigma - 2M^2_R)-t-{\nu}_t}\ ;
\label{1.11}
\end{equation}
\begin{equation}
M(s,t,u) =
 \sum\limits_{i=0}^{\infty}{\alpha}^{(u)}_i(u) {\nu}_u^i +
 \sum\limits_{R_s}^{}
      \frac{\rho^{(us)}(u)}{(\sigma - 2M^2_R)-u+{\nu}_u} +
 \sum\limits_{R_t}^{}
      \frac{\rho^{(ut)}(u)}{(\sigma - 2M^2_R)-u-{\nu}_u}\ .
\label{1.12}
\end{equation}
Here
$\sigma$
stands for the sum of squares of the external particle masses. The
right hand sides of the expressions
(\ref{1.10}) -- (\ref{1.12})
are written in terms of the natural coordinate systems in the
corresponding layers
\begin{equation}
B_x \{ x \in {\bf R},\ x \sim 0;\
    {\nu}_x \in {\bf C},\ |{\nu}_x| < \infty \}\ ,\\\\\
    x = (s,t,u) .
\label{1.13}
\end{equation}
This form of notations is convenient for the constructive
formulating of the analyticity conditions. Notice, that each pair of
layers
(\ref{1.13})
has a nonempty intersection:
\begin{equation}
D_s = B_t \cap B_u\ , \ \ \ \ \
D_t = B_u \cap B_s\ , \ \ \ \ \
D_u = B_s \cap B_t\
\label{1.14}
\end{equation}
(for example,
$t,u \sim 0$ in $D_s$, etc.).
Hence the summability conditions in every layer should be adjusted
in such a way that in the domain
\begin{equation}
R = B_s \cup B_t  \cup B_u
\label{1.15}
\end{equation}
they define the unique meromorphic function. The system of bootstrap
equations is an algebraic form of these matching conditions.

Taking into account the quoted above general considerations we
formulate the analyticity conditions for the amplitudes of binary
processes as follows%
\footnote{Let us emphasize that here we only discuss the scalar
amplitudes (the functions
$F^{(a)}$
in
(\ref{1.5}))}
\cite{AVVV}, \cite{AV}.
{\em The tree level amplitudes must be meromorphic functions in each
pair  energy
$s_{ij} \in {\bf C}$
at arbitrary fixed value of the second independent variable. In every
layer}
(\ref{1.13}),
{\em  containing the zero value hyperplane of one of the momentum
transfers
$x$,
they must be polynomially bounded functions of the corresponding
variable
${\nu}_x$.
The bounding polynomial degree
$N$
may depend on the quantum numbers characterizing the process.}

This formulation of analyticity conditions might seem unnecessarily
complicated, especially, if one takes into account that the domain
(\ref{1.15})
is only a part of the full complex space of two variables describing
the process.
We use it because of two reasons. First, the results of the papers
\cite{VV}, \cite{AVVV}
and
\cite{AV}
show that it leads to reasonable physical consequences. Second, even
in the case under consideration (binary processes) the corresponding
systems of bootstrap equations for the minimal parameters of the
extended theory turn out to be very complicated. In this case an
imprudent attempt to formulate more general requirements without
sufficient physical and mathematical motivation could lead to
inconsistency.

The examples that we analyze in the following sections illustrate the
structure and techniques of derivation of bootstrap equations.
However, before starting their consideration we would like to make a
short review of the Cauchy form method.

\section{The Cauchy forms}
\label{sec-forms}
\mbox{}

We are going to use the method (known from the complex analysis; see,
e.g.,
\cite{Shabat}), which allows one to present the polynomialy bounded
meromorphic function of one complex variable as a uniformly converging
series of pole contributions (in what follows we call such
representations as the Cauchy forms or Cauchy expansions). The
possibility to work in the layer
\begin{equation}
B_x \{ x \in {\bf R},\ x \in (a,b);\
       z \in {\bf C},\ |z| < \infty  \}
\label{1.16}
\end{equation}
(not only in the plane
$x = const$)
is provided by the natural modification of the method
(see \cite{VV}, \cite{AVVV}):
all the coefficients are considered to be smooth (real-analytic)
functions of the parameter
$x$.

First of all we need to specify the definition of the bounding
polynomial degree --- this turns out to be important for the analysis
of effective theories. We suppose that the reader is familiar with the
notion of the system of contours
$C_n$,
that appear in the definition of the polynomially bounded meromorphic
function of one complex variable
(see \cite{Shabat}).
When we work in the layer
(\ref{1.16})
we assume the smooth dependence of this system on the parameter
$x$.
The meromorfic function  $f(x,z)$   defined in the layer
(\ref{1.16})
is called as polynomially bounded with the degree
$N$
(or, simply
$N$-bounded),
if
$N$ is the minimal integer such that for all
$ x \in (a,b) $
\begin{equation}
{ \left.
\frac{|f(x,z)|}{z^{N+1}}
\right|}_{z \in C_m}
\stackrel{m \to \infty }{\longrightarrow} 0\ .
\label{1.17}
\end{equation}

The Cauchy form allows one to present the
$N$-bounded
in the layer
(\ref{1.16})
function
$f(x,z)$
as the uniformly converging series of pole contributions.
In the case, most interesting for our further purposes,
when all the poles are simple and there is no pole at
$z=0$,
it looks as follows:
\begin{equation}
f(x,z)=
\sum_{n=0}^N \frac{1}{n!}f^{(n)}(x,0)z^n +
\sum_{i=1}^{\infty}
\left\{
\frac{r_i(x)}{z-p_i(x)} - h^{(N)}_i(x,z)
\right\} \;\; .
\label{1.18}
\end{equation}
Here
$p_i(x)$ and $r_i(x)$
stand for the position of
$i$-th
pole and the corresponding residue. The poles are numbered  such
that
$$
|p_i(x)| \leq |p_{i+1}(x)|\ .
$$
The correcting polynomials
$h^{(N)}_i(x,z)$,
ensuring the convergence of the series look as follows:
\begin{equation}
h^{(N)}_i(x,z) \equiv -\frac{r_i(x)}{p_i(x)}\,
\sum_{n=0}^N\, {\left[ \frac{z}{p_i(x)} \right]}^n\ 
\equiv \sum_{n=0}^{N} h_{i,n}(x) \, z^n .
\label{1.19}
\end{equation}

It is not difficult to show
\cite{AVVV}, that for the 
$N-$bounded function
$f(x,z)$, represented by
(\ref{1.18}) in the layer
(\ref{1.16}), certain ``collapsing'' conditions are valid: the                            
correcting polynomial degrees of order higher than                              
$N$ converge themselves to the values of appropriate derivatives: 
\begin{equation}
\sum_{i=1}^{\infty} h_{i, N+k}(x) =
\frac{1}{(N+k)!}f^{(N+k)}(x,0)\ ,  \ \ \ \ \
x \in (a,b)\, \ \ \ \ \ k=1,2, \ldots\ .
\label{1.20}
\end{equation}
So, if in                                                                       
(\ref{1.18}) one uses some                                                           
$M>N$ instead of                                                                
$N$ (thus                                                                       
eq.~\ref{1.17} holds), the Cauchy expansion is still correct but can 
be reduced to the one with                                                         
$N$: the superfluous degrees of correcting polynomials just cancel              
higher order terms in the first sum of                                          
(\ref{1.18}).                                                                        
These conditions help us to puzzle out the system of
bootstrap equations; the corresponding example is given below.

The form
(\ref{1.18})
is the main tool used in
\cite{AVVV}, \cite{AV}
to derive the bootstrap equations.

\section{Bootstrap equations: a simple example}
\label{sec-example}
\mbox{}

Let us consider the simple example to illuminate the general scheme
discussed in the previous sections. It will allow us to show
explicitly how to obtain the bootstrap equations for the parameters of
a rational function of two variables, restricted by the corresponding
analyticity conditions.  This example obtains an explicit solution and
makes the terminology more transparent.

Consider the rational function of two complex variables
$F(x,y)$.
Let's demand (this is an analog of analyticity requirements) that
in the layer
\begin{equation}
B_y \{ x \in {\bf C};\
y  \in {\bf R},\  y \in (-\eta , +\eta) \}
\label{2.1}
\end{equation}
it  has the single pole (in
$x$),
and in the layer
\begin{equation}
B_x \{ y \in {\bf C};\
x \in {\bf R},\  x \in (-\xi, +\xi)  \}
\label{2.2}
\end{equation}
-- also a single pole (in
$y$).
Asymptotics is considered to be decreasing in each layer
(in terms of section
(\ref{sec-forms})
this function is 0-bounded in each layer).
The question that we are trying to answer is:
what is the structure of the set of the essential parameters
describing this function?

In this case the essential parameters are just the coefficients
$f_{ij}$
of the expansion
\begin{equation}
F(x,y) = \sum\limits_{i,j=1}^{\infty} f_{ij} x^i y^j\ .
\label{2.3}
\end{equation}
The posed above question can be phrased in a more concrete way: how
many independent combinations can be fixed arbitrarily and what are
these combinations?  Or, in  terms of field theory: how many
independent renormalization prescriptions is it necessary to impose in
order to fix the amplitude
$F(x,y)$
in the unique way,
and what is the explicit form of those prescriptions?

In the layer
(\ref{2.1})
$F(x,y)$
can be represented as follows:
\begin{equation}
F(x,y)  =
\frac{\rho (y)}{x - \pi (y)}\ \ , \ \ \ \ \ \ \ \ (x,y) \in B_y\ .
\label{2.4}
\end{equation}
The functions
$\rho (y)$
and
$\pi (y)$
are considered to be smooth in the vicinity of the origin:
\begin{equation}
\pi (y)  = \sum \limits _{i=0}^{} {\pi}_i y^i\ , \ \ \ \ \ \ \
\rho (y) = \sum \limits _{i=0}^{} {\rho}_i y^i\ .
\label{2.5}
\end{equation}

By analogy, in the layer
(\ref{2.2}):
\begin{equation}
F(x,y)  =
\frac{r(x)}{y - p(x)}\ \ ,\ \ \ \ \ \ \ \ (x,y) \in B_x\ ,
\label{2.6}
\end{equation}
where
\begin{equation}
p(x) = \sum \limits _{i=0}^{} p_i x^i\ , \ \ \ \ \ \ \
r(x) = \sum \limits _{i=0}^{} r_i x^i\ .
\label{2.7}
\end{equation}

In the intersection domain
$ B_x \cap B_y \equiv D_{xy} $
we obtain:
\begin{equation}
\frac{r(x)}{y - p(x)} = \frac{\rho (y)}{x - \pi (y)}\; ,
\ \ \ \ \
(x,y) \in D_{xy}
\{x \in (-\xi, +\xi) ,\; y \in (-\eta , +\eta)\} \ .
\label{2.8}
\end{equation}
Substituting
(\ref{2.5})
and
(\ref{2.7})
into
(\ref{2.8}),
we obtain an infinite system of conditions on the coefficients
$p_k, r_k, {\pi}_k, {\rho}_k$:
\begin{equation}
r_{i+1} {\pi}_0 - p_{i+1} {\rho}_0 = r_i, \ \ \ \
{\rho}_{i+1} p_0 - {\pi}_{i+1} r_0 = {\rho}_i, \ \ \ \
r_{i+1} p_{j+1} = {\rho}_{i+1} {\pi}_{j+1}\ \ \ \ \
i,j=0,1,...
\label{2.9}
\end{equation}
This system provides an example of what is called in
\cite{AVVV}
as the bootstrap equations. Once solved, it permits to express
the parameters
$p_i, r_i$
in terms of
${\pi}_i, {\rho}_i$.
It also gives an answer to the question if it is
possible to carry out the analytic continuation
from one layer to another. This is an infinite system of equations
with respect to
$2 \times \infty$
(formal notation!) unknown parameters,
which we need to reexpress the function
$F(x,y)$
in the layer
(\ref{2.2})
in terms of the parameters  defining it in the layer
(\ref{2.1}).
In general, it is very difficult to find the solutions of such systems
and even
to show  solvability. Fortunately, in this simple example it turns out
possible to give the explicit form of the solution.   This exercise is
really
useful because it gives an idea of the ``power'' of bootstrap
restrictions.

After we rewrite
(\ref{2.8})
as
\begin{equation}
r(x) [x-\pi (y)] = \rho (y) [y- p(x)]\ \ ,
\label{2.10}
\end{equation}
take the derivatives
${\partial}_{xy}^2,$
and separate the variables, we obtain:
\begin{equation}
\frac{r^{\prime}(x)}{p^{\prime}(x)} =
\frac{{\rho}^{\prime}(y)}{{\pi}^{\prime}(y)} \equiv a\ ,
\label{2.11}
\end{equation}
where primes mean the derivative with respect to the corresponding
variable, and
$a$ is the separation parameter. From
(\ref{2.11})
we obtain
\begin{equation}
r(x) = a p(x) + b\ , \ \ \ \ \ \ \ \
\rho (y) = a \pi (y) + c\ ,
\label{2.12}
\end{equation}
where
$b$
and
$c$ --
new constants. Finally, substituting
(\ref{2.12})
into
(\ref{2.10})
and separating the variables once more, we find
\begin{equation}
p(x) = \frac{d-bx}{c+ax}\ \ , \ \ \ \ \ \ \ \ \ \
\pi (y) = \frac{d-cy}{b+ay}\ \ ,
\label{2.13}
\end{equation}
were
$d$ --
another separation parameter.
The formulae
(\ref{2.13})
together with
(\ref{2.12}), (\ref{2.4})
and
(\ref{2.6})
give the exhaustive solution to the problem in question (it is easy to
check that the exceptional cases provide us nothing). The important
property of this solution is that it contains only 4 arbitrary
parameters! This means that the infinite system
(\ref{2.9})
only turns out to be  consistent if the function
$F(x,y)$
defined in the layer
(\ref{2.1})
belongs to the four-parametric family
\begin{equation}
F(x,y) = \frac{ad+bc}{-d + axy + bx + cy}.\
\label{2.14}
\end{equation}
This is the only case when there exists the analytic continuation of
this function from
$B_y$
into
$B_x$
with the desired properties. It is clear that in this case this
continuation
is unique.

The direct analysis of the system
(\ref{2.9}) would lead to the same conclusion.
It turns out possible in this simple example.
Unfortunately, the
regular method of solving the infinite-dimension algebraic systems is
not
known, except several trivial cases.

With the help of
(\ref{2.14}),
one can express the essential parameters
$$
f_{ij} = f_{ij}(a,b,c,d)
$$
in terms of ``fundamental constants''
$(a,b,c,d)$.
Then one can choose four arbitrary coefficients
$f_k$ ($k=1,2,3,4$)%
\footnote{Or four arbitrary combinations.},
that allow the inversion
$$
a=a(f_1,...,f_4),\ \ \ \ \ldots ,\ \ \ \ d=d(f_1,...,f_4)\ ,
$$
and impose arbitrary ``renormalization conditions'' for these four
quantities. The renormalization of all other essential parameters
should
respect the conditions
(\ref{2.9}).

Thus, now we can answer the question posed in the beginning
of this section.
{\em To fix the amplitude}
$F(x,y)$
{\em uniquely }
{\em it is sufficient to impose four renormalization prescriptions
fixing the ``fundamental'' constants}
$a,b,c,d$.

This example explains the prudence with which we have formulated
the analyticity conditions in section
(\ref{sec-analyticity}).
If, in addition to these conditions, one would impose supplementary
analyticity conditions (for example, in the layer
$$
C_x \{ y \in {\bf C};\
x \in {\bf R},\  x \in (1-\xi, 1+\xi)  \}\ ,
$$
with arbitrary number of poles and arbitrary asymptotic behavior in
this layer) then, except the lucky chance, he would fall in a
contradiction.

It is interesting to note that if we modify the problem and demand
that the function
$F(x,y)$
has one pole in the layer
$B_x$,
as in the previous case, but is 1-bounded (in place of 0-bounded) in
this layer, we would obtain a solution that depends also on 4
parameters. This solution, however, will be found among the
exceptional cases.

\section{Cauchy forms for the string amplitude }
\label{sec-string}
\mbox{}

The example, considered  in section
(\ref{sec-example}),
was too simple, and the method that was applied to solve it could
hardly be useful in the case of effective theories where the number of
poles is infinite. In this section we will show how to obtain the
bootstrap conditions for the function with infinite number of poles.
For this we use the techniques of Cauchy forms. Of course, we are not
able to show the explicit solution of those conditions.  However, we
will show that even in the case when the function
$F(x,y)$
is given explicitly (i.e. the set of minimal parameters is known) the
bootstrap conditions can be used as a source of non-trivial relations
connecting these parameters with each other. This very property was
used in
\cite{AVVV}, \cite{AV}
to obtain the restrictions on the physical characteristics of
pion-kaon and pion-nucleon scattering processes.

As an illustrative example we have chosen the Euler
$B$-function
(or, to be more precise, the so-called Lovelace amplitude
\cite{Lovelace},
which differs by a factor). This choice is explained by several
reasons%
\footnote{We also discussed this function in a slightly different
context in
\cite{AVVV}, Sec.~4.}.
First of all it is easy to follow the details of
calculations, because all the necessary identities are widely known.
Second, though there is known a great number of summation
formulae for Pochhammer symbols,
we obtain (using a very simple and extremely elegant method)
an infinite sequence of identities that could hardly be deduced with
the help of traditional methods.
Third, Euler's
$B$-function plays an important role in dual models and in string
theory
(see
\cite{Frampton}).
That is why our choice is justified from the physical point of view.
Finally, the last argument in favor of our choice is that the
numerical test of the corresponding bootstrap relations
allows us to understand qualitatively the structure of the
criteria that are necessary to evaluate the rapidity of convergence.
This point becomes very
important when one tries  to compare various theoretical predictions
(sum rules) with the experimental
data.

Let us consider the simple (string-like) model for the scattering
amplitude
that is constructed   -- in accordance with idea of Veneziano
\cite{Veneziano} --
out of
$B$-function
without a tachyon:
\begin{equation}
A(s,t) = (-s-t)B(\frac{1}{2}-s,\frac{1}{2}-t) =
\frac{\Gamma(\frac{1}{2}-s)\Gamma(\frac{1}{2}-t)}{\Gamma(-s-t)}.
\label{1}
\end{equation}

It has the following specific points (hyperplanes)
$(m,n = 0,1,2,...)$:
\begin{itemize}
\item
Zero hyperplanes:
$
s+t=n.
$
\item
Pole hyperplanes in
$s$ ($t$
fixed,
$s+t \neq m$):
$
s=\frac{1}{2}+n.
$
\item
Pole hyperplanes in
$t$ ($s$
fixed,
$s+t \neq m$):
$
t=\frac{1}{2}+n.
$
\item
Three series of ambiguity points located at the intersections of the
zero hyperplanes with the hyperplanes of poles in any variable. They
have the
following coordinates%
\footnote{One can find the corresponding plot in
\cite{AVVV}.}:
\\
Series
$A^{++}$:
$
s=+\frac{2m+1}{2},\ t=+\frac{2n+1}{2}.
$ \\
Series
$A^{+-}$:
$
s=+\frac{2m+1}{2},\ t=-\frac{2n+1}{2},\ \ \ \ \
(\ m \geq n).
$ \\
Series
$A^{-+}$:
$
s=-\frac{2m+1}{2},\ t=+\frac{2n+1}{2},\ \ \ \ \
(\ m \leq n).
$
\end{itemize}

Let us consider the behavior of the amplitude
$A(s,t)$
in the layers
$B_t\{ t\in {\bf R},\ s\in {\bf C},\ |s| < \infty \}$
with
$t \neq k+1/2$,
were
$k$ --
integer. The only singularities of the amplitude in such layers are
the poles
in variable
$s$.

Notice that, starting from some
$n$,
there is always a zero between the two poles of
$A(s,t)$.
For the contours
$C_n$
on the complex plane
$s$
we have chosen the system of circles (with the center at the
coordinate
origin)
passing through  zeroes of the amplitude.
It could be shown that everywhere on this system of contours, except
the
narrow sector in the vicinity of the real positive axis, the amplitude
$A(s,t)$
has the Regge type asymptotics.
$
(\sim s^{\frac{1}{2}+t})
$.
In the vicinity of real axis the asymptotics is controlled by the
presence of zero.

In the terminology of
Sec.~\ref{sec-forms}
in the layers%
\footnote{When treating the example with
$B$-function
we  use the natural shortened notations for the layers.}
$$
B_t \{ t \in (n-1/2, n+1/2); \quad n=0,1,... \}
$$
the amplitude
$A(s,t)$
is the
$n$-bounded
function of the complex variable
$s$
and of one real parameter
$t$.

In the layers
$$
B_t \{ t \in (-n-1/2, -n+1/2); \quad  n=1,2,... \}
$$
it has a decreasing asymptotics.

Residues of
$A(s,t)$
at the poles in
$s$
are the same in all the layers
$B_t$,
because the pole positions do not depend on
$t$.
In the case when the point under consideration is not the ambiguity
one, we have
\begin{equation}
r_n(t) \equiv
Res_{s=n+\frac{1}{2}}
\frac{\Gamma(\frac{
1}{2}-s)\Gamma(\frac{1}{2}-t)}{\Gamma(-s-t)}
= \frac{1}{n!}(\frac{1}{2}+t)\cdots
(\frac{1}{2}+t+n)
\equiv \frac{1}{n!}\, {\left( t+\frac{1}{2} \right) }_{(n+1)},
\label{1a}
\end{equation}
were
${\left( t+\frac{1}{2} \right) }_{(n+1)}$
stands for the so-called Pochhammer symbol (shifted factorial).

For example, let us construct the Cauchy expansion of
$A(s,t)$
when
$t \in (-3/2,-1/2).$
In this layer
$A(s,t)$
grows not faster then
$s^{0-\epsilon},\ \epsilon > 0$,
and there is no need in correcting polynomials. Here the Cauchy
expansion looks as follows:
\begin{equation}
A(s,t)=
\sum_{n=0}^\infty
\frac{1}{n!}\
\frac{(t+\frac{1}{2})_{(n+1)}}{(s-n-\frac{1}{2})}\ ,
\ \ \ \ \ \ \  t \in (-3/2,-1/2).
\label{4}
\end{equation}
Notice, that because the asymptotic becomes ``softer'' at large
negative
$t$,
this expansion is also valid at every
$t< -1/2$
(except the values  corresponding to the coordinates of ambiguity
points
$ t = -(2k+1)/2$
$(k=0,1,...)$,
where the expansion makes no sense%
\footnote{In what follows we do not mention this condition.}).

In the layer
$ B_t \left\{  t \in (-1/2, 1/2) \right\} $
the amplitude
$A(s,t)$
grows slower than a linear function of
$s$,
and thus in our Cauchy expansion we have to  account for the
correcting polynomials of
$0-$th
degree. Thus we obtain the following expansion:
\begin{equation}
A(s,t)=
A(0,t)+
\sum_{n=0}^\infty
\frac{1}{n!}\,
\left(
\frac{(t+\frac{1}{2})_{(n+1)}}{s-n-\frac{1}{2}} +
\frac{(t+\frac{1}{2})_{(n+1)}}{n+\frac{1}{2}}
\right) ,\ \ \ \ \
t \in (-1/2, 1/2).
\label{5}
\end{equation}
The techniques of the Cauchy forms allows us to represent the
meromorphic function of two complex variables as a converging series
of pole (in one variable) contributions; the convergence being uniform
in both variables. This will give us a possibility to obtain two types
of conditions on
$A(s,t)$:
the collapse conditions of superfluous degrees of correcting
polynomials (see                                                                            
(\ref{1.20})), and the bootstrap equations.

\section{Collapse conditions and bootstrap \\ for Pochhammer symbols. 
}
\label{sec-pohhammer}
\mbox{}

The collapse conditions 
(\ref{1.20}) on the regular part of the amplitude appear
when we pass from the layer where the amplitude has an increasing
asymptotics to another one, where the asymptotic regime is weaker.
The expansion
(\ref{5})
for
$A(s,t)$
in the layer
$B_t \{ t \in (-1/2,1/2) \}$
is also valid for
$t<-1/2$.
When
$t$
crosses the boundary value
$t=-1/2$,
corresponding to the change of asymptotic regime, the series of
correcting polynomials
can be summed independently:
$$
\sum_{n=0}^{\infty} h^{[0]}_n(t)=
\sum_{n=0}^\infty
\frac{1}{n!}\,
\left(
\frac{(t+\frac{1}{2})_{(n+1)}}{n+\frac{1}{2}}
\right) =
-\frac{\Gamma(\frac{1}{2}-t)\Gamma(\frac{1}{2})}{\Gamma(-t)} =
-A(0,t),\ \ \ \ \ t<-\frac{1}{2}\ ,
$$
and  the expansion
(\ref{5})
coincides with
(\ref{4}).

The bootstrap equations arise naturally from the requirement that the
Cauchy  expansion in one variable in some layer should coincide with
the expansion  in the cross-conjugated variable (i.e. in the
perpendicular layer) in the domain of intersection of these two
layers. For example, the expansion
(\ref{5})
is valid for the amplitude
$A(s,t)$
in the layer
$B_t \{ t \in (-1/2,1/2) \}$.
A similar expansion can be written in the layer
$B_s \{ s \in (-1/2,1/2) \}$:
\begin{equation}
A(s,t)=A(s,0)+
\sum_{n=0}^\infty
\left(
\frac{\rho_n(s)}{t-n-\frac{1}{2}}+\frac{\rho_n(s)}{n+\frac{1}{2}}
\right)\ ,
\label{6}
\end{equation}
where
$
\rho_n(s) \equiv \frac{1}{n!}
{\left( s + \frac{1}{2} \right)}_{(n+1)}
$.
These two expansions should coincide in the square formed by the
intersection of two layers. Hence in the domain
$(s \sim0, t \sim0)$
the following condition must be valid:
\begin{equation}
A(0,t) = A(s,0) + \sum_{n=0}^\infty
\left(
\frac{\rho_n(s)}{t-n-\frac{1}{2}}+\frac{\rho_n(s)}{n+\frac{1}{2}}
\right)
-\sum_{n=0}^\infty
\left(
\frac{r_n(t)}{s-n-\frac{1}{2}}+\frac{r_n(t)}{n+\frac{1}{2}}
\right)   \equiv
A(s,0)+\Psi(s,t).
\label{7}
\end{equation}

In some vicinity of the point
$(0,0)$
the function
$\Psi(s,t)$
is analytic because the corresponding series  converge uniformly; so
it is completely determined by the coefficients of its Taylor
expansion at this point. Let us differentiate both parts of the
equation
(\ref{7})
with respect to
$t$:
$$
\frac{\partial A(0,t)}{\partial t} =
\frac{\partial\Psi(s,t)}{\partial t}, \ \ \ \ \
(s \sim 0,\ t \sim 0)\ .
$$
The left hand side of this equality only depends on one variable
$t$.
This means that the dependence of the right hand side on the second
variable is purely fictitious. So one can assign to
$s$
any arbitrary value from the domain
$s \sim 0$
to compute the
$
\frac{\partial\Psi(s,t)}{\partial t} \; .
$
This allows us to determine the regular part of the amplitude
up to one arbitrary constant
$A(0,0)$.

These considerations allow us to rewrite
(\ref{7})
in the form of two conditions on the regular part of the amplitude
plus an infinite system of consistency conditions:
\begin{equation}
\frac{\partial A(0,t)}{\partial t} =
\frac{\partial\Psi(s,t)}{\partial t}\big|_{s=0},\ \ \ \ \ \ \
(t \sim 0)
\label{8}
\end{equation}
\begin{equation}
\frac{\partial A(s,0)}{\partial s} =
-\frac{\partial\Psi(s,t)}{\partial s}\big|_{t=0}, \ \ \ \ \ \ \
(s \sim 0)
\label{9}
\end{equation}
\begin{equation}
\frac{\partial^{k+p+2}}{\partial s^{k+1}\partial
t^{p+1}}\Psi(s,t)\big|_{s=0 , t=0}=0, \ \ \ \ \ \ \
\forall \  k,p=0,1,... \ .
\label{10}
\end{equation}
The consistency conditions   express  the fact that, in some vicinity
of the
point
$(0,0)$,
the derivative of
$\Psi(s,t)$
with respect to any variable does not depend on the cross-conjugated
variable.

Notice, that in this example the full symmetry between the variables
$s$
and
$t$
allows us to limit our analysis of consistency conditions to the case
$k > p$.

Such systems of conditions are called as bootstrap equations. They
represent nontrivial relations between the resonance parameters
(pole positions and residue values) of the function under
consideration. In the present example the pole position does not
depend on the cross-channel variable. In this case the system of
bootstrap equations leads to  an infinite set of relations for the
values of residues (Pochhammer symbols).

For example, let us consider the identity for the Pochhammer symbols,
following from
(\ref{10})
with
$k=1, p=0$:
\begin{equation}
\left\{
\sum_{n=0}^\infty
\frac{(-1)\ \rho_n^{(2)}(s)}{(t-n-\frac{1}{2})^2} -
\sum_{n=0}^\infty
\frac{(-1)^2\ 2!\ r_n^{(1)}(s)}{(s-n-\frac{1}{2})^3}
\right\}_{s=0,\ t=0} = 0.
\label{11}
\end{equation}
One can easily show, that the following equalities are valid for the
arbitrary order
derivative
of the residue:
$$
r_n^{(p)}(t) = 0,\ \ \ \ \ \ \ \ (p > n+1);
$$
$$
r_n^{(p)}(t)=
\underbrace{\sum_{i_1=0}^n\sum_{i_2=0}^{i_1-1}....
\sum_{i_p=0}^{i_{p-1}-1}}_{i_1>i_2...>i_p}
\frac{p!\ r_n(t)}{(\frac{1}{2}+i_1+t)...(\frac{1}{2}+i_p+t)},\ \ \
(p \leq n+1).
$$
This allows us to rewrite
(\ref{11})
in the following way:
\begin{equation}
\sum_{n=0}^\infty
\left\{
\frac{1}{(n+\frac{1}{2})^3}
\sum_{i_1=0}^n
\frac{1}{n!}\ \frac{(\frac{1}{2})_{(n+1)}}{(\frac{1}{2}+i_1)}
\right\}
-\sum_{n=1}^\infty
\left\{
\frac{1}{(n+\frac{1}{2})^2}
\underbrace{
\sum_{j_1=0}^n
\sum_{j_2=0}^{j_1-1}
}_{j_1>j_2}
\frac{1}{n!}\
\frac{(\frac{1}{2})_{(n+1)}}{(\frac{1}{2}+j_1)(\frac{1}{2}+j_2)}
\right\} = 0\ .
\label{12}
\end{equation}
One can observe, that already the first consistency condition provides
us with a highly non-trivial identity for the Pochhammer symbols. The
subsequent conditions lead to even more complicated identities. These
identities mirror the special properties of residues, which ensure the
existence of the solution of the bootstrap system. And we know it
exist - the solution is the Pochhammer symbols themselves, that is why
the relations above hold.

\section{Bootstrap system is overdetermined}
\label{sec-overfull}
\mbox{}

It is evident that the system of bootstrap equations is most probably
badly overdetermined. And the question that arises immediately is: how
to pick out a full subsystem or to tell what equations are evidently
unnecessary (related)?

In addition to the system of bootstrap equations  we have at our
disposal the collapse conditions of superfluous correcting polynomials
in the layers with softer asymptotic behavior. Now we will show how
with the help of collapse conditions and bootstrap in one layer it is
possible to obtain some of the bootstrap conditions in another layer.

In the three intersecting layers:
$ B_t \{ t \in (-1/2, 1/2) \} $,
$ B_t \{ t \in (-3/2,-1/2) \} $
and
$ B_s \{ s \in (-3/2,-1/2) \} $
the following Cauchy-expansions for the amplitude
$A(s,t)$
are valid:
\begin{equation}
A(s,t)=A(0,t)+
\sum_{n=0}^\infty
\left(
\frac{r_n(t)}{s-n-\frac{1}{2}}+\frac{r_n(t)}{n+\frac{1}{2}}
\right)\ , \ \ \ \ \
B_t \{ t \in (-1/2,1/2) \} ;
\label{a1}
\end{equation}

\begin{equation}
A(s,t)=\sum_{n=0}^\infty \frac{r_n(t)}{s-n-\frac{1}{2}}\ ,
\ \ \ \ \ \ \ B_t \{ t \in (-3/2,-1/2) \} ;
\label{a2}
\end{equation}

\begin{equation}
A(s,t)=\sum_{n=0}^\infty  \frac{\rho_n(s)}{t-n-\frac{1}{2}}\ ,
\ \ \ \ \ \ \ B_s \{ s \in (-3/2,-1/2) \} .
\label{a3}
\end{equation}

We demand that the corresponding expansions in the intersection
domains of each of two layers
$B_t$
with the layer
$B_s$
should represent the same meromorphic function. This allows us to
obtain the bootstrap conditions
\begin{equation}
A(0,t) =
\sum_{n=0}^\infty
 \frac{\rho_n(s)}{t-n-\frac{1}{2}}-\sum_{n=0}^\infty
\left(
\frac{r_n(t)}{s-n-\frac{1}{2}}+\frac{r_n(t)}{n+\frac{1}{2}}
\right)
\equiv \Psi_1(s,t),\ \ \ \ \ \ (s \sim -1,\  t \sim 0).
\label{14}
\end{equation}

\begin{equation}
0 = \sum_{n=0}^\infty
 \frac{\rho_n(s)}{t-n-\frac{1}{2}}-\sum_{n=0}^\infty
 \frac{r_n(t)}{s-n-\frac{1}{2}}
\equiv \Psi_2(s,t),\ \ \ \ \ \ (s \sim -1,\  t \sim -1).
\label{15}
\end{equation}

Using the same argumentation as in the previous section, we rewrite
(\ref{14})
as the system of the following conditions:
$$
A(0,t)=\Psi_1(-1,t),
$$
$$
{\left.
\frac{\partial^{k+p+1}}{\partial s^{k+1}\partial t^p}\Psi_1(s,t)
\right|}_{s=-1, \ t=0} = 0,\ \ \ \ \
\forall \ k,p=0,1,... \ .
$$
The first one of them gives the explicit expression for the regular
part of the
amplitude, while the second mirrors the independence of
$\Psi_1(s,t)$
of the argument
$s$.

The condition
(\ref{15})
could be rewritten in the similar form:
\begin{equation}
{\left.
\frac{\partial^{k+p}}{\partial s^{k}\partial
t^p}\Psi_2(s,t)
\right|}_{s=-1, \ t=-1} =0,\ \ \ \ \ \forall \ k,p=0,1,... \ .
\label{16}
\end{equation}

The explicit expression for
$A(0,t)$,
which is valid in the vicinity of
$t=0$,
could be without any obstruction analytically
continued to the domain of negative
$t$,
where the corresponding series converge very well.

The expansion
(\ref{a1})
is also valid at
$t<-\frac{1}{2} $.
~From the requirement that in this case it should coincide with
(\ref{a2}),
we obtain the collapse condition:
$$
A(0,t) =-\sum_{n=0}^\infty
 \frac{r_n(t)}{n+\frac{1}{2}}, \quad t<-\frac{1}{2}\ .
$$
This expression for the regular part of the amplitude should coincide
with the one obtained from
(\ref{14})
by the analytic continuation to the domain
$t<-\frac{1}{2}\ .$
In particular this  means that
$$
\Psi_1(-1,t)=-\sum_{n=0}^\infty
 \frac{r_n(t)}{n+\frac{1}{2}}\ ,
\ \ \ \ \ \  (s \sim -1,\ t \sim -1).
$$
As usual, let us rewrite this as a condition on the coefficients of
power series expansion
in the vicinity of
$(s=-1,\  t=-1)$:
$$
{\left.
\frac{\partial^{p}}{\partial
t^p}\Psi_1(-1,t)
\right| }_{t=-1} =
{\left.
-\sum_{n=0}^\infty \frac{r_n^{(p)}(t)}{n+\frac{1}{2}}
\right| }_{t=-1}.
$$
Using the explicit expression for
${\Psi}_1$,
we obtain:
$$
{\left\{
\sum_{n=0}^\infty
\frac{p!\ (-1)^p\rho_n(-1)}{(t-n+\frac{1}{2})^{p+1}}-
\sum_{n=0}^\infty
\left(
\frac{r_n^{(p)}(t)}{-1-n-\frac{1}{2}}
+\frac{r_n^{(p)}(t)}{n+\frac{1}{2}}
\right)
\right\}
}_{t=-1} =
{\left.
-\sum_{n=0}^\infty
 \frac{r_n^{(p)}(t)}{n+\frac{1}{2}}
\right|
}_{t=-1}.
$$
After collecting the similar terms one will find out that this
condition
coincides with
(\ref{16})
for all
$p$
when
$k=0$.

Thus the example of three layers  strengthens our confidence that the
system consisting of bootstrap equations and collapse conditions is
overdetermined. A share of information about bootstrap in lower layers
is contained in the bootstrap equations in upper layers and also in
the collapse conditions for transitions from upper to the lower
layers.

\section{Numerical test of the convergence rapidity}
\label{sec-numbers}
\mbox{}

The numerical test of the convergence rapidity of
the series
(\ref{12}) could be of great interest. This is because to check the
theoretical predictions of the dual models one often saturates the
quite similar to
(\ref{12})
identities (the so-called sum rules) with a
{\em finite}
number of resonances.

Unfortunately the up-to-date information on the hadron spectrum is far
from being exhaustive (especially in the region
$M > 2$ GeV).
This means that only those sum rules that converge sufficiently
rapidly could undergo the experimental  verification. That is why it
would be extremely instructive to learn how to pick these identities
out of the infinite system of bootstrap and collapse conditions. In
this section, by way of treating the example of the string amplitude
discussed in the
Sec.~\ref{sec-string},
we suggest a possible approach to this problem in the realistic
situation.

Let us carry out the numerical test of the system of identities for
residues
$r_n(t)$
and
$\rho_n(s)$,
obtained from the consistency conditions
(\ref{10}):
\begin{equation}
\left\{
\sum_{n=0}^\infty
\frac{(-1)^{p+1}(p+1)!\ \rho_n^{(k+1)}(s)}{(t-n-\frac{1}{2})^{p+2}}
\right\}_{{s=0} \atop {t=0}}-
\left\{
\sum_{n=0}^\infty
\frac{(-1)^{k+1}(k+1)!\ r_n^{(p+1)}(s)}{(s-n-\frac{1}{2})^{k+2}}
\right\}_{{s=0} \atop {t=0}} = 0,
\label{41}
\end{equation}
where
$k, p =0,1...,\  k>p$.
The condition
$k>p$
originates from the symmetry of spectrum in
$s$
and
$t$.

First of all we need to define a quantity that would allow us to
characterize the precision of saturation of the
sum rule
(\ref{41})
after one takes into account the finite number of items.
This could be done in the standard way, but in the current example
the procedure of calculation could be sufficiently facilitated.
However this point needs some comments  because
this is not always possible  in the realistic situations encountered
in the field theory.

In the expression
(\ref{41})
we deal with the difference of two absolutely convergent
numerical series.
Taking into account the  symmetry of spectrum, it looks natural
to consider the difference of the contributions
from the
$t$-
and
$s$-channel
poles at every step of the computation.
For the few first poles this contribution has
the definite sign (positive in the case
$k>p$)
but, starting from some number
$N_+(k,p)$,
depending on
$k$
and
$p$,
the sign of this difference changes.
Thus the convergence of this series to zero is provided
by the negative contribution of the large number of distant poles.
It  compensates gradually the positive
contribution of the first few poles.
As the convergence characteristics we chose the ratio:
$$
D \equiv \frac{\Delta S(N)}{S_+}\ ,
$$
where
$\Delta S(N)$
is  the discrepancy that remains after one considers
$2N$
poles ($N$ in the $s$-channel and $N$ in the $t$-channel),
and
$S_+$ --
is the sum {\em of all}
positive contributions (that correspond to the finite
number of initial terms of the series of differences)%
\footnote{Of course it only makes sense for
$N > N_+$.}.
With the help of
$D$
we can describe the convergence rapidity of the series
(\ref{41}).

It is sufficient to consider a small number of poles to reduce
significantly the relative discrepancy in the rapidly converging sum
rules.

The dependence of the relative discrepancy on the number of poles
taken into account for three first sum rules from the system
(\ref{41})
is shown on the picture.

\epsfig{figure=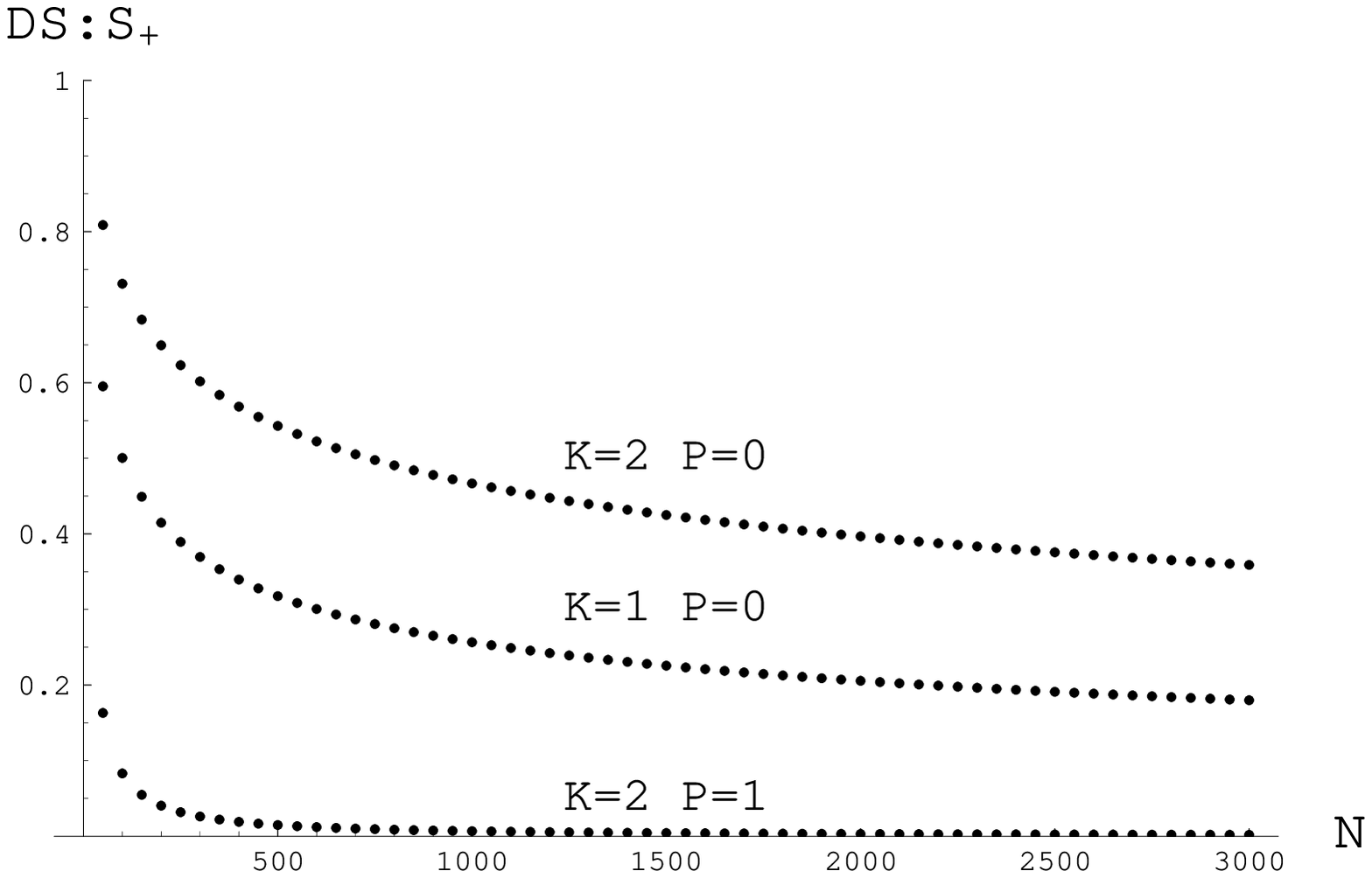, height=8cm}

The sum rule with
$k=2, p=1$
converges sufficiently fast. After one takes into account 100 first
poles%
\footnote{In each channel.}
the relative discrepancy equals approximately
8\%
(1\%
accuracy could be attained after accounting for 700 poles). Thus from
this condition we could obtain a sufficiently good relation between
the residues in the first 100 poles. The identities with
$k=1, p=0$
and
$k=2, p=0$
do not suite for this purpose. Sum rule with
$k=1, p=0$
converges much slower: the consideration of 3000 terms gives
$18\%$ discrepancy;
and to reduce it to
$9\%$,
one needs to take account of more than 22000 terms. The sum rule with
$k=2, p=0$
converges even more slowly.

The sum rules with best convergence are those
with
$p=k-1$
at large values of
$k$.
One can expect that for large
$k$
these sum rules would be saturated  rapidly.

It should be taken into account that
$
r_n^{(p)}(t) = 0
$
for
$p > n+1 $,
hence the first poles do not contribute.  Thus these sum rules could
serve as a source of relatively precise relations between the
parameters
of several resonances with
$n > p$.

The considered example allows one to understand in a qualitative way
the main properties
of the constructions arising from the bootstrap conditions.
In the realistic situation only the parameters of the few lightest
resonances are known.
The more astonishing
is that, as shown in
\cite{VV}, \cite{AVVV}
and
\cite{AV},
some of  the bootstrap restrictions are well saturated  by the
available experimental
data and provide the theoretical explanation to some phenomenological
relations.
This circumstance leads to the idea that the
$B$-function
gives a reasonable description only for the ``tail'' of the resonance
spectrum.
The parameters of the lower states are governed mostly
by the dynamical properties such as chiral symmetry.

\section{Conclusion}
\label{sec-conclusion}
\mbox{}

The examples considered above show
that the method of Cauchy forms is a useful tool
for deriving the relations between the parameters
of the polynomially bounded  meromorphic functions of two complex
variables.
It is easy to understand that, after the corresponding formulation
of the analyticity conditions,  this
method could be in principle applied to
the case of more variables.
But the practical advantage
of this approach to the study of inelastic amplitudes could hardly be
notable. Even in the case of
the simplest inelastic process
$2 \rightarrow 3$
one needs 5  independent variables. This leads to extremely bulky
expressions. The more powerful techniques is necessary that would
allow us to compactify the notations.

The case of binary processes is interesting because it allows one to
obtain the relations between the spectrum parameters following from
the correctness requirements of the perturbative scheme of the
$S$-matrix
calculation. As we already mentioned, many of those relations happen
to be in excellent agreement with the experimental data. This shows
that even such a complicated construction as effective field theory
could be successfully applied to the data analysis.

\section*{Acknowledgements}
\mbox{}

We are grateful to A.~Andrianov, A.~Vasiliev, M.~Vyazovski,
M.~Polyakov, V.~Sukhanov, H.~Nielsen, V.~Cheianov and J.~Schechter for
fruitful discussions of various problems associated with the concept
of the effective field theory. This work was supported by INTAS
(2000, project 587), RFBR (grant 01-02-17152), Ministry of Education
of Russia (E00-3.3-208) by the program ``Universities of Russia''
(UR.02.01.001) and Meltzers H\o yskolefond, Studentprosjektstipend
2002.


\end{document}